\documentclass[12pt,draft]{article}

\usepackage{amsthm,amssymb,amsmath,amscd,amstext,mathrsfs}
\usepackage{latexsym,mathptmx,tikz,caption}

\newtheorem{lemma}{Lemma}[section]

\newcommand{\nc}{\newcommand} 
\nc{\ca}{{\mathscr A}}

\nc{\N}{{\mathbb N}}
\nc{\Z}{{\mathbb Z}}
\nc{\T}{{\mathbb T}}
\nc{\R}{{\mathbb R}}
\nc{\C}{{\mathbb C}}
\nc{\HH}{{\mathbb H}}

\begin{document}

\title{On the emergence of the Lorentzian metric structure of space-time 
in general relativity}

\author{G\'abor Etesi\\
\small{{\it Department of Algebra and Geometry, Institute of Mathematics,}}\\
\small{{\it Budapest University of Technology and Economics,}}\\
\small{{\it M\H uegyetem rkp. 3., H-1111 Budapest, Hungary}}
\footnote{E-mail: {\tt etesi@math.bme.hu, etesigabor@gmail.com}}}

\maketitle

\pagestyle{myheadings}
\markright{G. Etesi: On the emergence of the Lorentzian structure}

\thispagestyle{empty}

\begin{abstract} In this short note we argue that, even if, as sometimes 
remarked, a Lorentzian manifold does not model correctly the structure 
of the continuum of physical events as it is, yet a Lorentzian 
manifold should describe its macroscopic structure as we experience it.

More precisely, theoretically motivated by von Weizs\"acker's 
chronological relative frequency interpretation of probability, and 
taking the Diaconis--Mosteller principle (also called the law of truly 
large numbers) as an empirical evidence in the macroscopic world, we 
argue that large collections of physical events appear in a composition 
of two fundamentally different formations, termed as progression and 
sample here, suggesting, in this framework, to use a 
Lorentzian-type metric on a manifold to describe matter-filled macroscopic 
regions of the physical continuum. 
\end{abstract}

\centerline{PACS numbers: 01.55.+b; 04.20.Cv; 05.50.+q; 11.30.Cp}
\centerline{Keywords: {\it Probability; Diaconis--Mosteller principle; 
Lorentz metric}}


\section{Introduction}
\label{one}


The {\it theory of general relativity} is an esteemed intellectual 
achievement within contemporary science, both \ae sthetically and 
scientifically. For this theory is not only beautiful 
thanks to its conceptual simplicity combined with amazing predictive 
power and mathematical elegance, it is also in a century-old perfect 
agreement with independent physical experiences, extending from 
accurate terrestrial laboratory and satellite experiments towards 
sophisticated deep space astronomical observations.

Apparently Einstein's theory correctly captures our impressions about 
time and space at meso- and macroscopic scales, including their 
structural similarity composed with their experiential difference. This 
co-existing similarity-difference of time and space enters the general 
relativity formalism as a fragile and tensionful balance between 
structures. It is declared that time and space exist in general 
relativity and they together comprise an $1+3=4$ dimensional physical 
continuum: the {\it space-time}, mathematically modeled by a pair 
$(M,g)$ consisting of a $4$ dimensional {\it differentiable manifold} 
$M$, which grasps the similarity by its homogeneity, and an $(1,3)$-type 
or {\it Lorentzian metric} $g$ on it, which certainly grasps the 
difference by its unusual signature. However $(M,g)$ is subject to 
Einstein's field equations which together with the theory's diffeomorphism 
invariance disturbs this simple division, for general relativity 
in this way becomes an intrinsically timeless theory, known as the 
``problem of time'' (cf. \cite[Appendix E]{wal} for a technical 
exposition, and {\it e.g.} \cite[Part 3]{bar}, \cite[Chapter 2]{cal} for 
broader physical and philosophical surveys). As a result, at a deeper 
layer of the theory the balance is lost rendering the role played by the 
Lorentzian signature of the metric less clear.

Apart from this, there has been a long and diverse debate in the 
scientific community concerning the nature and relevance of the 
co-existing similarity-difference of time and space. In the standard 
physics literature the emphasis is mainly put on similarity (except 
perhaps in classical thermodynamics and quantum measurement theory) in a 
well-known systematic way; while the much more complicated problem of 
difference has been approached from various directions (cf. {\it e.g.} 
\cite{bar,ber,cal,con-rov,hug-wut,hus,lib,rov,wei}) with yet only partial 
success. Nevertheless these efforts tend to reach at least one 
common insight, namely that difference between time and space gets 
more-and-more significant as the number of effectively available degrees 
of freedom in a physical system increases. This not necessarily means 
being macroscopic, as for instance the quite different roles played by time in 
large thermodynamic systems and in celestial mechanics indicates.

Returning to the mathematical formalization of general relativity, let 
us make a (certainly incomplete) list of problems, comments, questions, 
either physical or mathematical in their characters, motivated by the 
basic assumption of general relativity, namely that the physical 
continuum is modeled mathematically by a Lorentzian $4$-manifold 
$(M,g)$. Appearance of the mathematical continuum $M$ already alone 
does import serious difficulties into the 
general relativity discussions \cite{bae,ete,wey}, however let 
us rather focus here on puzzles stemming from the utilization of a 
Lorentzian-type metric $g$. In what follows we list items extending from 
physical towards mathematical issues, however doing so we do not mean an 
ordering by relevance. In our opinion, when treating difficulties in a 
physical theory, both conceptual-physical and technical-mathematical 
problems have an equal importance.

1. {\it Physical interpretation of Lorentzian geometric concepts in 
vacuum.} The vacuum Einstein equation with or without cosmological 
constant, admits non-trivial {\it i.e.}, curved solutions too. An example is 
the Schwarzschild geometry describing a static and spherically symmetric 
configuration of pure gravity {\it i.e.}, not shaped by any form of 
gravitating matter, rather forced solely by the unavoidable interaction 
of the gravitational field with itself. (Mathematically the 
self-interaction of gravity is reflected in the inherent non-linearity 
of the Einstein equation.) However even in this temporally static and 
spatially empty situation for example, the space-time is furnished with 
additional structures according to the {\it Lorentzian paradigm}. The 
most important extra structure is the collection of light cones (closely 
related to the causal structure). Geometrically, the two light cones 
intersecting in a common vertex at a space-time point emerge by tracing 
all null curves in space-time arriving at or departing from that 
space-time point. But what is the physical description of these curves? 
Following \cite{ehl-pir-sch} one of the most basic assumptions of 
general relativity is that space-time is a union of its points and these 
represent ``elementary physical events'', and accordingly the 
corresponding light cones are physically realized by ``light beams'' 
absorbed or emitted at these elementary physical events. But thinking 
physically, what kind of physical happening can occur in a space-time 
point in the completely static vacuous physical situation modeled by the 
Schwarzschild geometry? In particular, how light beams can be absorbed 
and emitted at these points or exist here at all {\it i.e.}, in an overall 
physical situation described by the static-empty Schwarzschild geometry? 
Taking the position of a strict operationalist (which however might be 
too narrow), light cones should appear only in those space-times which 
contain electromagnetic radiation, like {\it e.g.} the Kerr--Newman 
space-time, or at least some other propagating massless field. (Perhaps 
they can be introduced in space-times describing pure gravitational 
radiation too, but in this case some care is needed because the operational 
characterization of light cones looks self-referential.)

2. {\it Problems with the zero distance between distinct physical 
events.} Thinking in plain geometric terms the two light cones having 
a common vertex at a space-time point consist of those {\it distinct} 
space-time points whose Lorentzian distances from the given point are precisely 
zero. However thinking physically again, this is a strongly counter-intuitive 
feature of the {\it Lorentzian paradigm}. As Cornelius Lanczos puts it 
in his 1974 book on Einstein \cite[p. 32]{lan2}:

\begin{quotation}
\begin{small}

[...] Accepting the paradoxical situation concerning the `zero 
distance', we must be consistent and accept the leading principle of the 
Gauss--Riemann type of geometry, called `metrical geometry', that the 
distance is {\it the} quantity which decides all other geometrical 
properties. Now let us consider the following physical situation. It 
seems reasonable to assume that `zero distance' means strong physical 
interaction. We now consider an atom emitting light in the Andromeda 
Nebula. The light reaches Earth after three million years and interacts 
with the observing eye at the end of the telescope. However, 
the {\it four dimensional distance between the atom and the eye was zero 
throughout the three million years of travel}. Why then is the 
interaction restricted to the end of the journey, when the decisive 
geometrical quantity remains all the time zero, without any change?

I had a chance to discuss this question with Einstein who admitted its 
seriousness and felt very uncomfortable with an indefinite line element. 
Yes, he said, the indefinite metric offers a great puzzle which must 
arise from some deep seated source. But for the time being he did not 
see a solution and was willing to take the difficulty temporarily in his 
stride.
\end{small}
\end{quotation}

\noindent Working with general relativity nowadays, we already 
live in comfort with this fact and consider it as a subtlety or oddity of 
the widely accepted mathematical formulation of general relativity. But Lanczos 
continues with a warning \cite[pp. 32-33]{lan2}:

\begin{quotation}
\begin{small}

It seems to me that if we accept an apparent irrationality on empirical 
grounds, we will have to pay penalty sooner or later in heavy currency, 
as we had to pay the price for Newton's `absolute space' and `absolute 
time', and the equivalence of heavy and inertial mass. If we encounter 
an apparent irrationality at the very beginning of our speculations, we 
must feel permanently on dangerous ground.

\end{small} 
\end{quotation}

\noindent As a result of this critical attitude Lanczos worked out his own 
approach to the problem of indefiniteness, using classical field 
theoretic arguments \cite{lan1,lan3}: he supposes that the so-called 
classical vacuum is rather a composition of highly excited states and the 
originally Riemannian metric of the space-time continuum switches to a 
Lorentzian one only by averaging over these excited states, and the 
Lorentzian signature perhaps appears at macroscopic levels only.

3. {\it Ellipticity lost.} Partial differential equations (PDE's) are 
classified broadly as elliptic, parabolic and hyperbolic, exhibiting 
completely different behaviours and accordingly, requiring very 
different mathematical tools to handle them. For instance elliptic 
problems possess a very pleasant property called {\it elliptic 
regularity} leading, at least in the linear case, to a satisfactory 
theory regarding existence, uniqueness and differentiability of their 
solutions. Another extraordinary important tool in dealing with linear 
elliptic problems is the {\it Atiyah--Singer index theorem} connecting 
analysis and topology at a deep mathematical level. The most 
relevant fundamental differential equations of physics (like the Dirac, 
the wave, the gauge-fixed Yang--Mills, the linearized Einstein 
equations, {\it etc.}) are elliptic in Riemannian signature however are 
hyperbolic in Lorentzian one. Consequently, solving PDE problems in 
Lorentzian signature is often much less-tractable simply by lacking powerful 
mathematical techniques; which is in sharp contrast to the Riemannian 
case. Perhaps this is the right place to mention the various other 
mathematical difficulties arising from Lorentzian signature, but not 
necessarily related with ellipticity (like the non-compactness of the 
orthogonal group, extra difficulties in path integration, {\it etc., etc.}), 
too.

4. {\it Bad category.} It is attributed to the French mathematician 
Alexander Grothendieck the saying that ``It is better to have a good 
category with bad objects than a bad category with good objects.'' 
Whatever it precisely means, this is a summary of mathematicians' old 
experience that only those mathematical things are comfortable to work 
with, of whose category admits completions with respect to category 
theoretic (co)limits, by adding only ``reasonably singular'' limiting 
{\it i.e.}, ideal objects to the original ones. Most of compact manifolds 
cannot carry Lorentzian metrics (the obstruction is the manifold's Euler 
characteristic \cite[Theorem 40.13]{ste} which is not trivial in even 
dimensions) and this fact, together with the failure of the {\it 
Hopf--Rinow theorem} here (cf. \cite[p. 4]{bee-ehr-eas} makes the theory 
of {\it global Lorentzian geometry} quite vulnerable against 
counterexamples. Indeed, since most of Lorentzian manifolds are by 
default not complete, all kind of counterexamples can easily be 
constructed by artificially cutting out subsets (for an excellent 
summary see \cite{bee-ehr-eas}). However one has the permanent feeling 
that most of these counterexamples are somehow irrelevant from a 
``natural'' or ``physical'' point of view. This ``openness'' towards 
counterexamples indicates that the category of Lorentzian manifolds 
cannot be completed to a good category in the above sense; or considering its 
objects only, no good topology exists on the set of all 
Lorentzian manifolds (cf. {\it e.g.} \cite{cur,fle1,fle2,man}). 
In sharp contrast to this 
situation, the category of compact Riemannian manifolds is a good 
category because compactness is not an obstacle at all concerning 
Riemannian metrics hence the Hopf--Rinow theorem guarantees completeness 
of individual examples; consequently counterexamples are not so easy to 
find. Moreover, there is a satisfactory general limit construction here 
too, namely the {\it Gromov--Hausdorff convergence} of 
compact Riemannian manifolds.

5. {\it The Witt algebra.} From the abstract perspective of the general 
mathematical theory of finitely generated modules carrying inner products 
over a fixed ring \cite{mil-hus}, definite inner product spaces are 
distinguished in the following sense. Considering the relevant case of 
$\R$ here only, one can introduce certain equivalence classes of 
finite dimensionsl real vector spaces $V$ carrying scalar products 
{\it i.e.}, non-degenerate symmetric $\R$-bilinear forms 
$g:V\times V\rightarrow\R$. The 
tensor-product-over-$\R$ and the orthogonal-sum operations descend to these 
so-called Witt classes yielding a very important commutative ring with 
unit, the {\it Witt ring} $W(\R)$ of the reals (for a precise definition 
cf. \cite[Definition I.7.1]{mil-hus}). The signature of a scalar product 
is well-defined in every Witt class and in fact (see \cite[Corollary 
III.2.7]{mil-hus}) it gives rise to an isomorphism $W(\R)\cong\Z$. 
However one can prove (see \cite[Theorem III.1.7]{mil-hus}) that every 
element of the Witt ring over $\R$ {\it i.e.}, every Witt class of real scalar 
product spaces, contains a {\it canonical} representative, namely the 
unique space in this class which carries a {\it definite} scalar 
product. Consequently the aforementioned isomorphism arises simply by 
computing the signature of the unique definite scalar product in each 
Witt class whose absolute value is therefore equal to the dimension of 
the underlying real vector space. Putting differently, up to a natural 
equivalence in the theory of inner product spaces, in a given dimension 
there exist only two scalar product spaces over the reals, namely the 
ones which carry fully positive or negative definite scalar products.

By these (and probably furthers, to be added by the Reader) 
heterogeneous remarks one might feel uncomfortable with Lorentzian 
geometry and ask whether or not a Lorentzian manifold only a 
pseudo-structure in physics is. That is, perhaps it does not model any 
fundamental feature of the pure physical continuum shaped by 
gravity alone. However, even if this is the case, Lorentzian structure 
can still be relevant as an emergent macroscopic phenomenon, caused by 
the ``contamination'' of the pure physical continuum with 
macroscopic matter, which actually the real macroscopic physical 
situation is. In the remaining part of this note we make an attempt to 
bring this idea to a more solid basis, by evoking an interesting 
principle of applied statistics, and use it as an empirical argument to 
support the utilization of Lorentzian geometry in space-time physics at 
least at macroscopic levels.


\section{A conceptual analysis of the Diaconis--Mosteller principle}
\label{two}


The world surrounding us, apparently, perhaps as a manifestation of the 
{\it ergodic hypothesis}, has the tendency to realize the even most 
incredible but physically allowed situations. Studying the exciting and 
diverse problem of occurrences of surprising events, extraordinary 
coincidences in daily life, Persi Diaconis and Frederick Mosteller 
introduced a remarkable principle in 1989, termed as the {\it law of 
truly large numbers}, which in its full generality can be formulated as 
follows \cite[p. 859]{dia-mos}: 
\vspace{0.1in}

\centerline{\it With a large enough sample, any outrageous thing is likely to 
happen.}
\vspace{0.1in}

\noindent This principle offers a simple rational (but perhaps 
emotionally not satisfying) explanation how unlikely things can come to 
existence in the macroscopic world. (An analogous physical 
experience in the microscopic world is captured in Gell-Mann's {\it 
totalitarian principle}.) Our goal is to draw some physical conclusions from 
this principle; however before that we have to clarify some terms in its 
formulation.

The first of these terms is an {\it outrageous thing} what we 
immediately translate to the more neutral language of science as a {\it 
low probability physical event}, and then ask ourselves what {\it 
probability} is. In order to keep our exposition straight, we do not 
dive into this bottomless topic \cite{haj} here, rather directly proceed 
from von Weizs\"acker's chronological approach \cite{wei} (and defer 
some clarifications to Section \ref{four}), that 
probability is a mathematical way to model an obvious aspect of {\it 
time}, namely that the future consists of {\it possibilities} hence is 
objectively not known in the present (on the contrary, the past consists 
of {\it facts} hence is known). Consequently in general one can make 
only {\it predictions} on {\it forthcoming} physical events. We 
therefore accept that probability is the {\it predefined relative 
frequency} and in this way probability is an objective aspect of reality 
(to the extent to which physical time with its properties is objective). 
Note the subtlety that the chronological approach is free from the usual 
{\it circulus vitiosus} in the plain frequentist definition, for it 
supposes that probability can (somehow) be determined, in principle 
unambiguously in advance, and also supposes that any particular 
empirical sequence of relative frequencies in the corresponding 
truncated series of already-happened-physical-events, will converge to 
this known abstract real number; however, as a price, the exact 
convergence never can be confirmed in any finite experiment. We also 
accept that the mathematical formalization of classical probability is 
Kolmogorov's theory. Now we can make our 
\vspace{0.1in}

\noindent {\bf Assumption 1.} (Existence and uniqueness of probability) 
{\it Probability, as the limit of a sequence of relative frequencies in 
a progression of independent physical copies of a model event, is a 
well-defined scalar taking values in $[0,1]\subset\R$. That is, this 
real number exists and is independent of the particular observer who 
records the events, makes unbiased judgments about their favourability, and 
then computes relative frequencies and their limit in the progression.} 
\vspace{0.1in}
 
\noindent Here by a {\it progression} we mean an at most countable set 
$E$ of physical events, which for an observer $O$ appear in an 
ordering $<_O$\:, might be called as ``strictly later than'', 
according to the observer's proper time ({\it i.e.}, simultaneous detection 
of events is not permitted). Then, as usual, the {\it relative 
frequency} of the detected and accordingly ordered truncated set of events 
$e_1<_Oe_2<_O\dots<_Oe_n$ taken from $E$ is the ratio 
$\frac{\vert\mbox{favourable cases}\vert}{n}\in [0,1]$. Although for an 
objectively given $E$ the ordering of events, the judgments on their 
favourability, the values of the corresponding relative frequencies, 
{\it etc., etc.}, can depend on the particular observer at finite $n$'s, it is 
assumed that whenever $n\rightarrow+\infty$ the limit exists 
unambiguously and independently of any particular {\it ordinary} 
observer $O$, giving rise to the probability $p\in[0,1]$ which is therefore 
unambiguously assigned to $E$ ({\it i.e.}, {\it not} to an individual event 
$e$).

Next we turn to the {\it likely to happen} term in the above 
daily formulation of the Diaconis--Mosteller principle. The precise 
formulation of this idea in terms of probability leads to our

\vspace{0.1in}

\noindent {\bf Assumption 2.} (Diaconis--Mosteller principle) {\it The 
probability that an event occurs within a sample of independent physical 
copies of a model event having positive probability, gets arbitrarily 
close to $1\in[0,1]$ as the finite cardinality of the sample increases.}

\vspace{0.1in}

\noindent Finally by the term {\it sample} we {\it a fortiori} mean 
precisely that sort of collection of physical events for which the 
formulated probabilistic behaviour, as an empirical evidence for an {\it 
ordinary} observer $O$, holds. However note that in this way two similar 
concepts enter our discussion: the ``progression'' and the 
``sample'' of physical events. Both of them certainly label specific 
collections of physical happenings in {\bf Assumption 1} and {\bf 
Assumption 2} respectively, however their distinction is not clear. The 
following simple but key technical lemma demonstrates that they are in 
fact strictly different, hence validating the terminology:

\begin{lemma} Suppose that {\bf Assumption 1} and {\bf Assumption 2} are 
valid. Take a collection $\{e_1,\dots,e_k\}$ consisting of $0<k<+\infty$ 
independent physical events with uniform probability $0<p<1$. Provided 
the cardinality $k$ is large enough, $\{e_1,\dots,e_k\}$ is 
strictly either a progression or a sample, if any. 
\label{kulcslemma} 
\end{lemma}

\begin{proof} Since the complementers of independent events are also 
independent, the probability that an event with probability $0\leqq p\leqq1$ 
{\it does not} happen in $k$ independent trials is $(1-p)^k$ 
hence the probability that it does happen in the collection 
$\{e_1,\dots,e_k\}$ is $q=1-(1-p)^k$. Suppose {\bf Assumption 2} 
holds and $\{e_1,\dots,e_k\}$ is a sample and $0<p\leqq 1$; then 
$q\rightarrow 1$ as $k\rightarrow+\infty$. Suppose that {\bf Assumption 1} 
also holds and $\{e_1,\dots,e_k\}$ is also a progression and $0<p<1$ is 
the relative frequency of favourable events; then $q\approx 1$ for large $k$ 
implies that in this case in fact $p\approx 1/k$ that is, there exists a 
constant $0<c<+\infty$ independent of $k$ satisfying $0<p\leqq c/k<1$ for 
sufficiently large $k$'s. However in this case 
\[q=1-(1-p)^k\leqq 1-\left(1-\frac{c}{k}\right)^k
\leqq 1-\frac{{\rm e}^{-c}}{2}<1\] 
for large $k$'s contradicting that $q\rightarrow 1$ as 
$k\rightarrow+\infty$.  Thus a sufficiently large collection with 
uniform probability $p\not=0,1$ cannot be both a progression and a sample. 
\end{proof}

\noindent Consequently Lemma \ref{kulcslemma} forces us to 
distinguish progressions and samples from each other whenever $0<p<1$. 
One can argue that, as 
an aspect of macroscopic reality, large ensembles of physical events 
arrange themselves phenomenologically for an {\it ordinary} observer in two 
fundamentally different ways. One is characterized by 
{\bf Assumption 1}, in which an ensemble exhibits a more-and-more sharp 
but otherwise arbitrary probability $p\in(0,1)$ as its size 
increases, hence the ensemble approaches a purely probabilistic 
description; this is what we call a {\it progression} (but also might be 
termed as a {\it timelike arrangement}). The other is characterized by 
{\bf Assumption 2}, in which all probabilities behave like $q\rightarrow 1$ 
as the size increases, hence the ensemble approaches a purely 
deterministic description; this is what we call a {\it sample} (but also 
might be termed as a {\it spacelike arrangement}). {\bf Assumption 1} also 
guarantees that this distinction is independent of 
the particular {\it ordinary} observer. The general pattern of a large 
ensemble is then a mixture of the two pure cases {\it i.e.}, a progression 
and a sample. In the extreme cases $p=0$ hence $q=0$ and $p=1$ hence $q=1$ 
the conclusion of Lemma \ref{kulcslemma} fails; hence 
in particular in the latter deterministic situation progressions and 
samples are indistinguishable. 

Before proceeding further we would like to make a comment. In our 
understanding the empirical evidence summarized in the 
Diaconis--Mosteller principle ({\bf Assumption 2} here) and the 
mathematical fact expressed in Lemma \ref{kulcslemma} together imply 
that probability and cardinality are {\it independent} data of a large 
ensemble of physical events. Moreover by accepting von Weizs\"acker's 
chronological relative frequency interpretation of probability 
\cite[Teil II.2]{wei} in {\bf Assumption 1}, we can 
also say that probability characterizes the functional or temporal 
distribution, while cardinality independently characterizes the 
structural or spatial distribution of a large ensemble of physical 
events. In this way our observation offers an argument on an empirical 
({\it i.e.}, not metaphysical, or mathematical, or theoretical physical, 
psychological, {\it etc}.) ground that in the continuum of physical events 
there exist ``mainly probabilistic'' or ``timelike'' direction(s) as 
well as ``mainly deterministic'' or ``spacelike'' direction(s) at least 
macroscopically. From this angle our approach is similar to that of 
Callender who argues for the existence of a distinguished timelike 
direction in the continuum of physical events. The only apparent 
difference is that contrary to the empirical argument here, 
\cite[Chapters 6-8]{cal} rather uses mathematical, especially 
algorithmic compressibility considerations.


\section{Recovering the Lorentzian structure}
\label{three}


As a next and plain technical step, we work out a mathematical model for 
the continuum of physical events which turns out to be nothing else than a 
Lorentzian manifold. This can be considered as an analogue of 
the geometric substantiation of space-time carried out {\it e.g.} in 
\cite{ehl-pir-sch} however goes along conceptually quite different lines. 

First consider a discrete picture. Let $\{e_1,\dots,e_m\}$ be 
a sample as in {\bf Assumption 2} consisting of $m$ independent physical 
events of equal {\it a priori} probability $0<p<1$. The probability $q$ 
that the event occurs in the sample thus satisfies $q\rightarrow 1$ 
as $m\rightarrow+\infty$. For every $e_i\in\{e_1,\dots,e_m\}$ in this 
sample put $e_{i,1}:=e_i$ and as in {\bf Assumption 1} let 
$\{e_{i,1},e_{i,2},\dots, e_{i,n}\}$ be one of the (in principle 
many) possible progressions for this event, whose relative frequency therefore 
converges to $p=p(e_{i,1})$ whenever $n\rightarrow+\infty$. Lemma 
\ref{kulcslemma} makes 
sure that $\{e_{1,1},\dots,e_{m,1}\}\not=\{e_{i,1},\dots,e_{i,n}\}$ 
for every $i=1,\dots,m$ {\it i.e.}, the sample differs from all progressions of 
its elements as well as by the independence of events obviously 
$\{e_{i_1,1},\dots,e_{i_1,n}\}\cap\{e_{i_2,1},\dots,e_{i_2,n}\}=\emptyset$ 
for every $1\leqq i_1<i_2\leqq m$ {\it i.e.}, the progressions are disjoint. 
In this way we roughly obtain an $m\times n$ grid 
$E_{m,n}:=\{e_{i,j}\}_{\begin{smallmatrix}i=1,\dots,m\\
                                 j=1,\dots ,n
              \end{smallmatrix}}$ of events. 

We would like to pass to the continuum limit, more precisely carry 
out this construction in a continuum of physical events too. To achieve 
this, first we recall some mathematical facts summarized in 
\cite[Section 4]{ete}. Consider a connected oriented compact, or more 
generally a connected oriented finite-volume Riemannian manifold $(M,g)$. 
This basic mathematical structure, conventionally considered as possessing a 
rigid geometric nature exhibiting lengths, angles, curvatures, {\it etc.} only, 
in fact canonically gives rise to a Kolmogorov probability measure space 
$\big(M,\ca_g, p_g\big)$, 
too. Indeed, the orientation of the manifold $M$ and the Riemannian metric 
$g$ on it together induce a measure $\mu_g$ on $M$; then define $\ca_g$ to be 
the $\sigma$-algebra consisting of all $\mu_g$-measurable subsets of $M$, 
including $M$; then writing $V_g:=\int_M\mu_g<+\infty$ for the volume of 
$(M,g)$ and putting $p_g:=\frac{1}{V_g}\mu_g$ we obtain a probability measure 
given by $p_g(A)=\frac{1}{V_g}\int_A\mu_g$ for $A\in\ca_g$. The remarkable 
point is that this probability space up to diffeomorphisms is independent of 
the particular metric $g$ used to construct it; for it follows from a 
theorem of Moser \cite{mos} that if $(M,h)$ is another similar Riemannian 
structure on $M$ yielding $(M,\ca_h,p_h)$ then there exists an 
orientation-preserving diffeomorphism $f_{h,g}:M\rightarrow M$ such that 
if $A\in\ca_g$ then $f_{h,g}(A)\in\ca_h$ and $p_h(f_{h,g}(A))=p_g(A)$. 
Thus we conclude that there is a unique 
({\it i.e.}, metric-independent) way to convert finite-volume Riemannian 
manifolds into probability measure spaces. We can think of the 
geometrically-given universal probability $p_g$ as for some $A\in\ca_g$ the 
number $0\leqq p_g(A)\leqq 1$ is the probability that an elementary physical 
event occurs in this prescribed measurable region {\it i.e.}, if $e\in M$ 
then in fact $e\in A$ (for a support of this interpretation 
cf. \cite[{\bf Assumption-physical form}]{ete}).

Proceeding now further, note that if $B\in\ca_g$ then 
$q_g(B)=\frac{1}{V_g}\int_B\mu_g\rightarrow 1$ whenever 
$\int_B\mu_g\rightarrow V_g$, resembling the probability-cardinality 
relationship in {\bf Assumption 2}. Given an {\it ordinary} observer 
$O$, we can therefore think of the Riemannian manifold $(M,g)$ as a 
non-countable analogue of a large sample having regularized infinite 
cardinality equal to its finite volume. In light of the above interpretation 
of the geometric probability, a point is always ``smudged'' over some 
measurable subset hence we can effectively identify $e\in M$ with its 
appropriately small neighbourhood $e\subseteqq U_e\subseteqq M$ and put 
$p(e):=p_g(U_e)$. 

However by {\bf Assumption 1} we have to assign a 
progression of events to every $e\in M$ too, such that the relative 
frequencies in this progression converge to the 
probabilities $0\leqq p(e)\leqq 1$ just introduced. From now on suppose that 
$e\subsetneqq U_e\subsetneqq M$ hence strictly $0<p(e)<1$. In this case by 
Lemma \ref{kulcslemma} the progression is not a subset of the sample $M$. 
Thus take another differentiable manifold $N$ and a map $\pi: N\rightarrow M$ 
and by exploiting the many options, suppose that 
the sought progression for $e$ is sufficiently regular in the sense that 
$\pi^{-1}(e)\subset N$ contains it (for all $e\in M$). Since $M$ is a 
manifold, we also suppose roughly that there exists a model 
differentiable manifold $F$ such that $\pi^{-1}(e)\cong F$ for every $e\in M$. 
Consider an {\it ordinary} observer $O_e$ at the event $e$. As a subtle but 
important point, notice that the demand on $\pi^{-1}(e)$ that as detected by 
$O_e$ it admits an {\it ordering} $<_{O_e}$, forces to choose, as an only 
option, $F\cong\R$ and globally we have to suppose that $N$ is oriented and 
$\pi:N\rightarrow M$ is orientation-preserving. We have accepted the 
effective identification of $e\in M$ with its neighbourhood 
$e\subsetneqq U_e\subsetneqq M$. Consequently we refine $\pi^{-1}(e)\cong\R$ 
and more precisely suppose that the local triviality 
property of vector bundles holds {\it i.e.}, 
$\pi^{-1}(U_e)\cong U_e\times\R$ such 
that $\pi^{-1}(x)\cong\R$ is a linear isomorphism for every $x\in U_e$. 
The uniqueness part of {\bf Assumption 1} {\it i.e.}, the assumption that 
probability is independent of a particular {\it ordinary} observer who 
records the corresponding relative frequencies, finally makes sure that 
the resulting structure of $N$ is independent of any particular {\it ordinary} 
observer who was used to construct it. In this way we can quite naturally 
conclude that $N$ has the structure of an oriented real line bundle that 
is, an oriented rank-$1$ real vector bundle over a manifold $M$. 
From now on we embed $M$ into $N$ by the zero section as usual. 

The discrete and the continuum pictures are unified by taking an 
inclusion $E_{m,n}\subset N$. We suppose that this inclusion is 
structure-preserving {\it i.e.}, the sample is mapped into $M$ while 
progressions into fibers of $\pi: N\rightarrow M$. More precisely, we 
suppose that $E_{m,n}\cap M=\{e_{1,1},\dots,e_{m,1}\}$ is the sample part of 
$E_{m,n}$ as well as for every $e_{i,1}\in E_{m,n}\cap M$ 
we suppose that $E_{m,n}\cap\pi^{-1}(e_{i,1})=\{e_{i,1},\dots,e_{i,n}\}$ 
is its progression part within $E_{m,n}$. We can think of the discrete subset 
$E_{m,n}$ as a collection of physical events distinguished by 
observation within an ambient continuum $N$ of physical events. Finally 
we make $E_{m,n}$ macroscopic by letting both $m\rightarrow+\infty$ and 
$n\rightarrow +\infty$ such that $E_{m,n}\subset N$ converges to an 
everywhere dense countable subset $E\subset N$.

As a next step, pick an event $e\in E\cap M\subset N$ and consider the 
corresponding tangent space $T_eN$. The fibre bundle structure $\pi: 
N\rightarrow M$ satisfying $\pi^{-1}(e)\cong\R$ induces $T_eN 
=T_eM\oplus T_e\pi^{-1}(e) \cong T_eM\oplus\R$ and recall that via the 
Riemannian structure $(M,g)$ there already exists a partially defined 
symmetric non-degenerate positive definite bilinear form 
$g_e:T_eM\times T_eM\rightarrow\R$. It is therefore natural to extend $g_e$ 
to a full symmetric non-degenerate 
bilinear form $h_e:T_eN\times T_eN\rightarrow\R$. By the 
Jacobi--Sylvester theorem \cite[Theorem III.2.5]{mil-hus} this extension is
uniquely characterized by the signature of $h_e$. This signature emerges from 
our assumptions made in Section \ref{two} precisely as follows.

\begin{lemma}
Suppose that {\bf Assumption 1} is valid, more precisely its 
ordinary-observer-in\-dependence-of-pro\-bability part holds true 
infinitesimally. Also suppose the validity of {\bf Assumption 2}. 
Then the signature of $h_e$ on the real vector space $T_eN$ of dimension 
$\dim_\R N=\dim_\R M+1$ is equal to $\dim_\R M-1$ i.e., $h_e$ is a Lorentzian 
scalar product.
\label{kiterjesztes}
\end{lemma}

\begin{proof} To find the signature of $h_e$, take smooth curves 
$\gamma,\delta:\R\rightarrow N$ satisfying 
$\gamma(0)=\delta(0)=e\in E\cap M$ such that, according to the 
{\it ordinary} observer $O_e$ as above, $\gamma$ connects events 
in $E\cap M$ belonging to the distinguished sample while $\delta$ connects 
events in $E\cap\pi^{-1}(e)$ belonging to the distinguished progression of 
$e$ (see the left side of Fig. 1). Referring to Lemma \ref{kulcslemma} which 
says that samples and progressions are necessarily different, as well as 
taking into account that $E\subset N$ is everywhere dense by construction, we 
can suppose that $\dot{\gamma}(0)$ and $\dot{\delta}(0)$ are linearly 
independent within $T_eN$ (see the right side of Fig. 1). 
\vspace{0.2in}

\centerline{
\begin{tikzpicture}[scale=0.7]
\node at (-10, 2) {$N\supseteqq V_e$};
\draw [thick] plot [tension=0.8,smooth] coordinates 
{(-8,1) (-7,2) (-6,0) (-5,1) (-4,2) (-3,2)};
\draw [fill=black] (-7,2) circle (0.15cm);
\draw [fill=black] (-6,0) circle (0.15cm);
\draw [fill=black] (-4,2) circle (0.15cm);
\node at (-8,0.4) {$\gamma$};
\draw [fill=black] (-5,1) circle (0.15cm);
\node at (-4.5,1) {$e$};
\draw [thick] plot [tension=0.8,smooth] coordinates 
{(-5.2,-1) (-5,1) (-5,2) (-4.5,3) (-4.8,4)};
\draw [fill=black] (-4.5,3) circle (0.15cm);
\node at (-4.3,4) {$\delta$};
\node at (1,2) {$T_eN$};
\node at (3.2,3.5) {$\dot{\delta}(0)$};
\draw [fill=black] (3,1) circle (0.15cm);
\draw [thick] [->] (3,1) -- (3.1,3);
\draw [thick] [->] (3,1) -- (4,1.7);
\node at (4.9,1.5) {$\dot{\gamma}(0)$};
\node at (3.5,1) {$e$}; 
\end{tikzpicture}}
\vspace{0.1in}

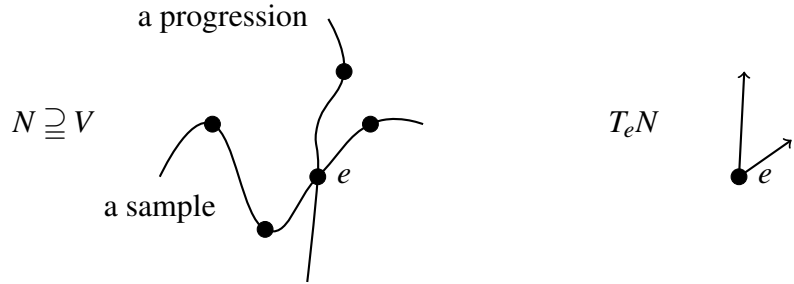
\centerline{Figure 1. The local (left) and the infinitesimal (right) shape 
of a large progression and a large sample}
\centerline{through an event $e$ in the ambient continuum $N$ of physical 
events.}
\vspace{0.2in}

\noindent Take an orientation-preserving diffeomorphism 
$f:N\rightarrow N$ such that $f(e)=e$ and the corresponding 
derivative $f_*(e):T_eN\rightarrow T_eN$ is an $h_e$-orthogonal 
transformation. Recalling the usual distinction between gauge and 
symmetry transformations\footnote{Given a space-time $(N,h)$ a general 
element $f\in{\rm Diff}^+(N)$ is a {\it gauge} transformation {\it i.e.}, 
$(N,h)$ and $(f(N),f^*h)$ are considered physically the same 
space-times; however if in addition $f^*h=h$ holds {\it i.e.}, $f\in{\rm 
Iso}^+(N,h)\subsetneqq {\rm Diff}^+(N)$ then $f$ is called a {\it 
symmetry} transformation {\it i.e.}, $(N,h)$ and $(f(N),h)$ are considered 
physically different, cf. {\it e.g.} \cite[p. 438]{wal}.} we can 
interpret $f$ as a {\it symmetry transformation} interchanging 
two observers $O'_e$ and $O''_e$ meeting at $e\in N$. The uniqueness part of 
{\bf Assumption 1}, namely that probability is an invariant scalar, implies 
that progressions and samples cannot be transformed into each other 
by $f$ globally; indeed, if $f(E\cap\pi^{-1}(e))\subseteqq E\cap M$ could occur 
then meanwhile $O'_e$ computes in the usual way that 
$p(e)=\lim\limits_{n\rightarrow+\infty}\frac{\big\vert\mbox{favourable cases in
$E_{m,n}\cap\pi^{-1}(e)$}\big\vert}{n}<1$ the other $O''_e$ would compute 
$p(e)=\lim\limits_{m\rightarrow+\infty}\frac{\big\vert\mbox{favourable cases in
$E_{m,n}\cap M$}\big\vert}{m}=1$ by {\bf Assumption 2}. Hence 
$f(E\cap\pi^{-1}(e))\not\subseteqq E\cap M$ and in particular 
$f(\delta)\not\subseteqq\gamma$. Accepting the uniqueness part of 
{\bf Assumption 1} infintesimally as well, we therefore get 
$f_*(e)\dot{\delta}(0)\not=c\dot{\gamma}(0)$ and likewise 
$f_*(e)\dot{\gamma}(0)\not=c\dot{\delta}(0)$ where $c\not=0$ is some constant. 
Thus the orbit of $\dot{\gamma}(0)$ under the group of $h_e$-orthogonal 
transformations cannot contain $\dot{\delta}(0)$ and {\it vice versa}, 
therefore the orthogonal group induced by $h_e$ does not act transitively 
on the rays of $T_eN$. Hence it follows at once that the non-degenerate 
symmetric bilinear form $h_e$ on $T_eN$ here is necessarily {\it indefinite}. 
Concerning its signature, we know already that $h_e\vert_{T_eM}=g_e$ is 
positive definite and that there exists an at least $1$ dimensional subspace 
of $T_eN$ on which $h_e$ is negative definite; however $T_eM\subset T_eN$ has 
codimension $1$ hence the negative definite subspace of $T_eN$ is at most $1$ 
dimensional too. Thus $h_e$ has Lorentzian signature. 
\end{proof}

\noindent Two comments are in order here. The first is that in our opinion 
Lemma \ref{kiterjesztes} does not have an intrinstic mathematical content 
{\it i.e.}, its strength depends only on the relevance of 
{\bf Assumption 1} (and concerning this point also cf. Section \ref{four}). 
The second is that Lemma \ref{kiterjesztes} fixes the signature of $h_e$ only 
and does not say anything on the {\it dimension} of $T_eN$ {\it i.e.}, of $N$. 
Nevertheless, repeating Lemma \ref{kiterjesztes} 
over all points of the everywhere dense subset $E\subset N$ and assuming 
smoothness we extend $h_e$ over $N$ and come up with a Lorentzian manifold 
$(N,h)$ as claimed. 

Consider finally the extreme case of $e\subsetneqq U_e=M$ hence 
$p(e)=p_g(M)=1$ for every $e\in M$. At the expense of completely 
eliminating spatial localizability of events, this is a fully 
deterministic situation. By the failure of Lemma \ref{kulcslemma} in 
this case, distinguishability does not hold hence progressions can be 
absorbed into samples; thus we can terminate our construction with the 
{\it Riemannian} space $(M,g)$ we began with. That is, in our framework 
the extension of the Riemannian $(M,g)$ to a Lorentzian $(N,h)$ is 
unnecessary in this rather unphysical situation. Nevertheless the 
inclusion of this extreme case in our discussion demonstrates that the 
existence of a large ensemble of events with non-trivial probabilities, 
or referring again to the chronological interpretation of probability, 
eventually the macroscopically existing physical time itself, is the key 
in the emergence of the Lorentzian signature (in our approach).

Based on our two plausible probability assumptions of Section \ref{two}, we 
have been able to equip the physical continuum fulfilled with macroscopic 
matter with the expected structure of a Lorentzian manifold, called a 
space-time in general relativity in the broad sense. More precisely, we 
could reproduce only the signature of the metric {\it i.e.}, its infinitesimal 
structure. The next question is whether or not can one go further 
and say something on the local and then the global structure of $h$ that 
is, reproduce the Einstein equation in this framework over $N$ too, and 
then obtain a space-time in the strict sense {\it i.e.}, a Lorentzian manifold 
satisfying a specific Einstein equation.\footnote{Note that the treatment of 
the Einstein equation in the substantiation of general relativity in 
\cite{ehl-pir-sch} is missing.} Concerning this important point, observe 
that our constructions so far suffer from an ambiguity. Given an event 
$e\in M\subset N$, on the one hand we have constructed its probability 
from the spatial metric $g$ along its sample $M$ as $p_g(U_e)$ where 
$U_e\subset M$ is an open subset about $e\in $M; on the other hand 
{\bf Assumption 1} says that the probability $p(e)$ of the same event is 
assigned to (one of) its progression(s) in $N$. We also know via Lemma 
\ref{kulcslemma} that 
the progression of $e$ cannot be a subset of $M\subset N$ hence thinking 
of $p_g(U_e)$ as the predefined relative frequency of the progression, the 
asserted equality $p_g(U_e)=p(e)$ above assumes some relationship between 
the 1 codimensional Riemannian submanifold $(M,g)$ and its ambient space 
$(N,h)$. We {\it conjecture} that this relationship is the Einstein 
equation.\footnote{Appearance of the Einstein equation, usually 
considered as having a highly deterministic character, in a 
probabilistic context is not surprising in light of the aforementioned 
canonical connection between finite-volume Riemannian manifolds and Kolmogorov 
probability spaces.} Although we have not been able (yet) to derive 
precisely the Einstein equation itself or any of its equivalent reformulations 
in this way, we can at least, as a first compatibility check, exhibit a 
promising evidence in this direction as follows. Introducing the extrinsic 
curvature $k$ of $(M,g)\subset (N,h)$ we can write 
the Einstein equation on $(N,h)$ with matter $T$, satisfying the dominant 
energy condition, in the form of constraint equations along 
$(M,g)\subset (N,h)$, cf. \cite[Equations 10.2.41 and 10.2.42]{wal}, as 
usual: 
\[\left\{\begin{array}{ll} s_g -\vert k\vert_g^2 +{\rm 
tr}^2k=16\pi\rho ,\\ 
{\rm div} (k-({\rm tr}\:k)g) =8\pi J, \\ 
\rho\geqq\vert J\vert_g\geqq 0. 
\end{array}\right.\] 
In the Riemannian $m$-manifold $(M,g)$ the small-radius-expansion of the 
volume of a ball of radius $r$ about a point $x\in M$ exists and looks like 
${\rm Vol}_g(B_x(r))=\frac{\pi^{\frac{m}{2}}} 
{\Gamma(\frac{m}{2}+1)}r^m\big(1-\frac{s_g(x)}{6(m+2)}r^2+\dots\big)$. 
Inserting only the first of the Einstein constraint equations ({\it i.e.}, the 
full Einstein equation as well as the dominant energy condition are not 
used), the local probability can therefore be written as 
\[p_g(B_e(r))=\frac{1}{V_g}\:\frac{\pi^{\frac{m}{2}}} 
{\Gamma(\frac{m}{2}+1)}r^m\left(1-\frac{16\pi\rho(e)- {\rm tr}^2k(e)+
\vert k(e)\vert^2_g}{6(m+2)}\:r^2+\dots\right)\] 
demonstrating that, in the presence of the Einstein constraints, 
$p_g(B_e(r))$ is {\it not} determined by the pointwise matter energy 
density $\rho(e)$ at $e\in M$ and the local spatial 
geometry about $e\in M$ through the radius $0<r<\varepsilon$ alone, but also 
depends on the local temporal geometry about $e\in N\supset M$ through the 
extrinsic curvature tensor $k$ (equal to the time derivative of $g$). 
Consequently writing $p_g(B_e(r))=p(e)$ is consistent, for $p(e)$ is also 
given by the relative frequency of a progression in $N$ as in 
{\bf Assumption 1} and probability has been interpreted chronologically 
i.e., progressions are timelike. 


\section{Conclusion and outlook: dicing around a black hole}
\label{four}


In the previous sections we have exhibited an argument to support the 
utilization of Lorentzian metrics at least in {\it ordinary} macroscopic 
situations in general relativity, despite the known physical and mathematical 
problems related with the inconvenient signature of these type of 
metrics. In this closing section we make couple of 
general remarks and clarifications. 

First let us make a comparison of our approach with the causal set 
theory program (cf. \cite{hug-wut} for a recent survey). The general aim 
of the latter program is to obtain the whole space-time structure of classical 
general relativity or even quantum gravity from very first principles 
such as elementary causal relations on plain sets. Our aim here is 
however less ambitious: upon accepting that physical happenings in {\it 
ordinary} macroscopic situations already appear in a continual structure 
mathematically modeled by a smooth manifold (cf. the {\it 
Hauptvermutung} in \cite[Subsection 4.1.4]{hug-wut}), we simply ask why 
Lorentzian and not Riemannian its geometric structure is? Therefore in 
our opinion the problem-setting here is closer to {\it e.g.} \cite{lan1,lan3} 
or \cite[Chapters 6-8]{cal} than to that summarized in \cite[Chapters 
3-4]{hug-wut}.

Second, we would like to give some clarifications on the choice of our 
probability interpretation here. Following \cite{haj}, existing 
probability interpretations so far can be roughly divided into four major 
branches: {\it classical, frequentist, subjective} and {\it propensity}. 
Although any of these interpretations might work well in appropriate 
contexts, taking into account the empirical-physical character of the 
considerations exhibited here, we feel that the first one as well as the 
last two of these interpretations are not really suitable for our 
purposes because of their too strong mathematical, 
psychological-cultural and metaphysical contents, respectively. Our 
choice, namely the von Weizs\"ackerian chronological interpretation of 
probability \cite[Teil II.2]{wei} belongs to the so-called {\it 
predictive frequentist} school; which means that probability is 
understood as the {\it right now} given limit of the relative frequency 
of {\it similar} physical events or experiments to happen or to be 
executed {\it in the future}. That is, in this approach in addition to 
the plain frequentist interpretation, not only an empirical sequence but 
the temporal character of this sequence {\it i.e.}, the existence of {\it 
physical time} is essential too. Consequently this probability 
interpretation is in natural harmony with the broad conviction that 
the unusual signature of the Lorentz metric in general relativity 
somehow reflects, at the mathematical level, the relevance of physical time 
in the theory (despite the known ``problem of time'' issues). In particular, 
as we mentioned already, our approach also fits into the scheme set up by 
Callender \cite[Chapters 6-8]{cal}.

Nevertheless our selected probability interpretation has its own 
intrinsic difficulties. For example, there indeed appears to be a 
certain tension in invoking this particular probability interpretation 
in the context of an empirical argument as this interpretation accepts 
that probability predictions never can be exactly confirmed empirically. 
Another headache is that speaking, when computing relative frequencies, 
of so-called {\it similar} physical events is problematic in a {\it 
chronological} context as these are supposed to be separate {\it 
individual} physical events in physical time, whose exact similarity or 
their relation to a ``model event'' is therefore ill-defined in general.

Third, we make a comment on the extendibility of our particular probability 
interpretation to more general {\it i.e.}, not only {\it ordinary} general 
relativistic situations. Throughout the text and especially around the 
formulation of {\bf Assumption 1} and {\bf Assumption 2} we have carefully 
emphasized that an {\it ordinary} observer's experiences are used to 
substantiate the whole set-up here. However accepting that general 
relativity describes not only our mild terrestrial experiences but works 
correctly in the dynamical and strong range of gravity too (which is 
supported by the recent direct gravitational wave and black hole 
observations) the important question arises how to interpret probability 
in these more general situations \cite{sau}. Recall that in our framework 
probability is the predefined relative frequency of a sequence of events 
(called progression here); this abstract real number then can be 
verified by an observer with certain accuracy (but never exactly) by 
detecting these events and counting the favourable cases. We have 
assumed that probability in this way is a well-defined scalar. 
Acknowledging this observer-dependence our worry can be put as follows: 
{\it to what extent is probability a relativistically invariant scalar 
quantity?} Before targeting this problem note that some of the aforementioned 
other approaches to probability are apparently more safe because they do 
not refer to any observer. Consider for instance the so-called {\it 
classical interpretation} by Laplace via symmetry considerations 
\cite{haj}: the probability of getting the number 6 in one toss with an 
unloaded dice is $\frac{1}{6}$ precisely because this dice is a {\it 
perfect cube} having 6 equal sides, {\it etc}. This elegant forecasting of 
probability is then indeed verified (with finite accuracy) in a sequence 
of events. However note that in this empirical verification process we 
have assumed by convention that the dice and the gamer more-or-less stay 
together in time and space what we call an {\it ordinary} situation. Instead of 
this imagine for example that the dice is on the board of a distant 
spaceship moving, with respect to the gamer, with a velocity comparable 
with the speed of light. Apparently the shape of the dice is distorted 
by Lorentz contraction hence the gamer concludes that the probability is 
not equal to $\frac{1}{6}$; meanwhile receiving a message with the 
result of a long sequence of trials, recorded by the ship's crew, will 
support $\frac{1}{6}$. Thus it seems the classical and the relative 
frequency interpretations of probability diverge already in a simple 
special relativistic situation. This conclusion is however wrong: in our 
context very surprising computations \cite{pen, ter} demonstrate that 
(as long as a single eye is used and the image occupies a small solid 
angle) the effect of Lorentz contraction on the visual appearance of a 
solid body is not distortion but rather a perfect rotation! Hence quite 
interestingly spatial symmetry is preserved at least in the realm of 
special relativity. However imagining more wild games (dicing around a 
neutron star, or in the vicinity of a rotating-oscillating black hole 
whose space-time geometry is even not known yet) one would not expect a 
similar ``conspiration'' which could save the coherence of various 
probability concepts.

Thus, as a no better option, we return to {\bf Assumption 1} and ask 
again to what extent probability an observer-in\-de\-pen\-dent scalar 
quantity is. Imagine the situation inspired by \cite{ete-nem}: consider 
two observers, one is the gamer orbiting on a stable equatorial orbit 
around a massive slowly rotating Kerr black hole while the other is a 
traveler holding an unloaded dice and departing into the black hole 
without hitting its central singularity (hence in principle can survive 
the trip). According to his own clock, the traveler will cross the 
(outer) event horizon in finite time and we can suppose that he starts 
tossing the dice strictly after this event only. Meanwhile the traveler 
already inside the black hole will likely conclude that the relative 
frequency approaches the usual $\frac{1}{6}$ probability (but honestly 
speaking: who knows?), the gamer outside the black hole and watching the 
game from a distance surely cannot reach a definitive answer ever. This 
bit awkward example indicates two things: despite the validity of {\bf 
Assumption 1} in usual cases, assigning a progression (from the many 
possible ones) hence a probability to a ``model event'' is not obvious, 
or even {\bf Assumption 1} itself might break down in truly generic 
relativistic situations. 

Apart from and independently of these worries, 
of course the validity of {\bf Assumption 2} gets also questionable in 
too extremal situations. However these conjectured limitations of our 
approach in inexperienced distant situations are consistent with the 
empirical character of these considerations.

\vspace{0.1in}

\noindent{\bf Acknowledgements}. The author is grateful to the Reviewers 
for the excellent questions and suggestions, which helped to clarify some 
parts of the arguments here.

\vspace{0.1in}

\noindent{\bf Declarations and statements}. No any form of AI was used 
to write this article. All not-referenced results in this work are fully 
the author's own contribution. There are no conflict of interest to 
declare that are relevant to the content of this article. The work meets 
all ethical standards applicable here. No funds, grants, or other 
financial supports were received. Data sharing is not applicable to this 
article as no datasets were generated or analyzed during the underlying 
study.

\end{document}